# MgBeC: A potential MgB$_2$-like Superconductor

O. P. Isikaku-Ironkwe[1, 2]
[1]The Center for Superconductivity Technologies (TCST)
Department of Physics,
Michael Okpara University of Agriculture, Umudike (MOUAU),
Umuahia, Abia State, Nigeria
and
[2]RTS Technologies, San Diego, CA 92122

## Abstract

Intensive and extensive searches for superconductivity in materials iso-structural, iso-electronic and iso-valent with magnesium diboride, MgB$_2$, have not produced superconductors with Tc close to 39K. Recently, we proposed lithium magnesium nitride, LiMgN, as a potential MgB$_2$-like superconductor, based on symmetry considerations of averages of electronegativity, valence electrons, atomic number and formula weight. Our studies show that there are other materials that meet the same symmetry considerations like LiMgN. Here we propose that magnesium beryllium carbide, MgBeC, one of such materials, should be a superconductor like MgB$_2$, with Tc near 39K.

## Introduction

Search for superconductivity in borides was a well-explored field in 1970, that ruled out exploring materials with valence electron count much below 3.8 [1], like magnesium diboride, MgB$_2$. It was therefore a surprise that MgB$_2$ was found to be superconducting [2] with Tc of 39K. Efforts both theoretical [3 - 7] and experimental [8 - 17], based on structural, electronic and electrovalent symmetry has failed to produce true magnesium diboride-like superconductors. Approaching the problem from symmetry of electronegativities, valence electrons and atomic numbers has opened a rich treasury of new possibilities of magnesium diboride analogues [18 - 22]. Recently [20], we showed that Lithium Magnesium nitride, LiMgN, meets the symmetry conditions first proposed in [19]. Here we introduce a new material, Magnesium Beryllium Carbide (MgBeC) that also meets the same symmetry conditions as LiMgN. This paper is structured as follows: we first describe the probable route to making MgBeC. Next we compare MgBeC with MgB$_2$ and LiMgN via their Material Specific

Characterization Datasets (MSCDs) [19] and then invoke the symmetry search rules for superconductivity and the Tc equation to determine the Tc of MgBeC.

**Preparation and Properties of MgBeC**
Though no records were found for preparation and properties of MgBeC, possible preparation routes may be:

$$Mg_2C + Be_2C = 2MgBeC \quad (1)$$

or:

$$0.6667 Mg_2C_3 + Be_2C + Be + 2.3333 Mg = 3 MgBeC \quad (2)$$

Beryllium carbide, $Be_2C$, and Magnesium carbide, $Mg_2C$, are well-studied ionic antifluorite semiconductors [23, 24, 25] that are isovalent and cubic. Magnesium sesquicarbide, $Mg_2C_3$, We guess that MgBeC may have a cubic structure too and may be semiconducting at room temperature.

**Predicting superconductivity and Tc in MgBeC**
MSCD analysis of MgBeC shown in Table 1, indicates that it has the same electronegativity, valence electrons and atomic number as $MgB_2$ and LiMgN and meet the same test for superconductivity, namely $0.75 < Ne/\sqrt{Z} < 1.02$. We have showed [19] that the maximum Tc of a material may be expressed in material specific parameters of electronegativity, $\mathcal{X}$, valence electrons, Ne, and atomic number, Z, given by

$$T_c = \mathcal{X} \frac{Ne}{\sqrt{Z}} K_o \quad (1)$$

where $K_o$ is a parameter that determines the value to Tc. $K_o = n(Fw/Z)$ and n is dependent on the family of superconductors. Fw represents formula weight of the superconductor. For $MgB_2$, $K_o = 22.85$ and Fw/Z is 6.26, making n = 3.65. Recently [20] we proposed that similar superconductors may arise when we compare their averages of electronegativity, $\mathcal{X}$, valence electrons, Ne, and atomic number, Z, and Fw/Z. We distinguished four possible cases, when at least two of the features are the same, namely: (a) ⟨$\mathcal{X}$, Ne, Z⟩, (b) ⟨Ne, Z⟩, (c) ⟨$\mathcal{X}$, Ne⟩ and (d) ⟨$\mathcal{X}$, Z⟩. We described such superconductors as similar. We found the symmetry rules apply within the range for most high-Tc superconductors. The symmetry rule, first proposed in [19] and applicable herein as shown in Table 1, is:

*"Materials with exactly the same average electronegativity, valence electrons and atomic number have the same Tc"*

The symmetry of their MSCDs gives us strong ground to predict that LiMgN will be a superconductor like $MgB_2$ but with Tc of 38.5K

## Discussion

The search for magnesium diboride-like superconductors was based on the symmetry of nature and the evidence even from nature that there are "families" so magnesium diboride cannot be "the only kid in town". In a previous paper [20] we proposed LiMgN as a potential $MgB_2$ superconductor. MgBeC meets exactly the same symmetry criteria. There are no reports that it has ever been produced. Following the rule that materials with the same valence electrons can be combined, we suggest that $Mg_2C$ and $Be_2C$ may combine to yield MgBeC. In a previous paper [20] we showed that iso-structural and iso-valent symmetry were insufficient to give the high Tc of $MgB_2$ in such similar compounds. Here we show again like in [20] the MSCD symmetry [19] are the key factors that lead to superconductivity and high Tc. Like for LiMgN, this claim awaits experimental verification.

## Conclusion

Symmetry considerations used to predict LIMgN as a superconductor, apply exactly to MgBeC. If MgBeC stoichiometry exists, we predict it will be found to be a superconductor with Tc of 38.5K, just like LiMgN.

## Acknowledgements

The search for $MgB_2$-like superconductors is an adventure story in mind connectivity involving two continents and many dramatis personae. Most significant are A.O.E. Animalu at University of Nigeria, Nsukka, M.B. Maple and J.E. Hirsch at UC San Diego and J.R. O'Brien at Quantum Design, San Diego. My special thanks to M. J. Schaffer for financial support.

# TABLE

| | Material | $\mathcal{X}$ | Ne | Z | Ne/$\sqrt{Z}$ | Fw | Fw/Z | Tc(K) | Ko |
|---|---|---|---|---|---|---|---|---|---|
| 1 | MgB$_2$ | 1.7333 | 2.667 | 7.3333 | 0.9847 | 45.93 | 6.263 | 39 | 22.85 |
| 2 | LiMgN | 1.7333 | 2.667 | 7.3333 | 0.9847 | 45.26 | 6.172 | 38.5 | 22.85 |
| 3 | MgBeC | 1.7333 | 2.667 | 7.3333 | 0.9847 | 45.33 | 6.1814 | 38.5 | 22.85 |

Table 1: Material Specific Characterization Dataset (MSCD) for MgB$_2$, LiMgN and MgBeC. MSCD helps us to see immediately the relationships between two or more materials in terms of electronegativity, $\mathcal{X}$, valence electrons, Ne, atomic number, Z, formula weight, Fw. The exact match for $\mathcal{X}$, Ne and Z and close match of their Formula weights (Fw), lends very strong grounds for predicting very close Tcs of these materials, following the chemical symmetry rules of ref. [19].

.